\documentclass[12pt,a4paper]{iopart}

\usepackage{iopams}
\usepackage{graphicx}

\newcommand{\ket}[1]{\mbox{$|#1\rangle$}}

\begin{document}

\title{Anti-symmetrisation reveals hidden entanglement}

\author{Alessandro Fedrizzi$^{1,3}$, Thomas Herbst$^1$, Markus Aspelmeyer$^1$, Marco Barbieri$^{3,4}$, Thomas Jennewein$^{1,3,5}$, Anton Zeilinger$^{1,2}$}

\address{$^1$Institute for Quantum Optics and Quantum
Information, Austrian Academy of Sciences, Boltzmanngasse 3, 1090
Wien, Austria}
\address{$^2$Faculty of Physics, University of Vienna, Boltzmanngasse 5, 1090
Vienna, Austria}
\address{$^3$Department of Physics and Centre for Quantum Computer
Technology, University of Queensland, QLD 4072, Australia}
\address{$^4$Groupe d'Optique Quantique, Laboratoire Ch. Fabry de l'Institut dÕOptique, F91127 Palaiseau, France}
\address{$^{5}$Institute for Quantum Computing, University of Waterloo, 200 University Avenue West, Ontario N2L 3G1, Canada}

\begin{abstract}
Two-photon anti-bunching at a beamsplitter is only possible if the photons are entangled in a specific state, anti-symmetric in the spatial modes. Thus, observation of anti-bunching is an indication of entanglement in a degree of freedom which might not be easily accessible in an experiment. We experimentally demonstrate this concept in the case of the interference of two frequency entangled photons with continuous frequency detunings. The principle of anti-symmetrisation of the spatial part of a wavefunction and subsequent detection of hidden entanglement via anti-bunching at a beamsplitter may facilitate the observation of entanglement in other systems, like atomic ensembles or Bose-Einstein condensates. The analogue for fermionic systems would be to observe bunching.
\end{abstract}

\maketitle

Entanglement represents the most striking departure of the quantum world from ordinary experience: the correlations exhibited by entangled particles are inexplicable with the concepts of classical physics. Such non-classical correlations power the advantages offered by quantum systems for computation and metrology. Entanglement is a precious resource which needs a careful characterisation, but, in some occasions, it can be difficult to distinguish from purely classical correlations. This is the case, for instance, for the frequency of single photons; while a classical energy correlation between two light quanta is relatively easy to check, assessing its quantum nature by observing correlation in a frequency superposition basis is technically extremely challenging. A strikingly simple solution is to study the parity of multi-particle states, since it can be affected by entanglement to the point that it may lead to an apparent inversion of the spin statistics. Entangled bosons can, under certain conditions, appear as fermions and vice versa.

Consider, as an example for a bosonic quantum system, single photons. If two separable and indistinguishable photons coincide on a symmetric beamsplitter (BS), they will, due to the bosonic commutation relation, leave the BS through the same output port \cite{hong1987mst}. However, in the presence of entanglement one can also observe the opposite; the fact that the singlet state of two polarization-entangled photons can be unambiguously identified via anti-bunching at a beamsplitter is a well-established method in quantum information, e.g. for Bell state measurements \cite{michler1996ibs}, specifically in dense coding \cite{mattle1996}, teleportation \cite{bouwmeester1997eqt} or in linear optical quantum computing \cite{kok2007loq}. This phenomenon is based on the fact that interference on a BS reveals the spatial symmetry of a wavefunction; whenever its spatial part is antisymmetric, bosons must anti-bunch \cite{wang2006qtt,eckstein2008bfm}.

Here we observe quantum beatings of frequency entangled photons generated in spontaneous parametric downconversion (SPDC). Our experimental scheme is novel in that it combines the well known technique of observing two-photon interference dips by varying the arrival times of photons at a beamsplitter \cite{hong1987mst} and the ability to continuously tune the frequencies of the involved photons by a change of a single experimental parameter. The frequency degree of freedom has only indirectly been manipulated in two-photon interference experiments so far, for example via discrete selection of frequencies with filters or apertures in spatial quantum beating experiments \cite{ou1988osq,rarity1990tcp,larchuk1993iep,saleh1998set,kim2003tpi}. 

By anti-symmetrising the wavefunction and observing quantum beatings we are able to detect the presence of entanglement, which would otherwise remain unobserved, without actually accessing the degree of freedom in question. Anti-bunching at a BS has been observed many times since the first quantum beating experiment \cite{ou1988osq}. We want to point out explicitly that the general precept for these observations was always either a priori spatial anti-symmetry or deliberate anti-symmetrisation, which is a method applicable to general quantum systems.

\begin{figure}[!b]
\begin{center}
\includegraphics[width=0.6\textwidth]{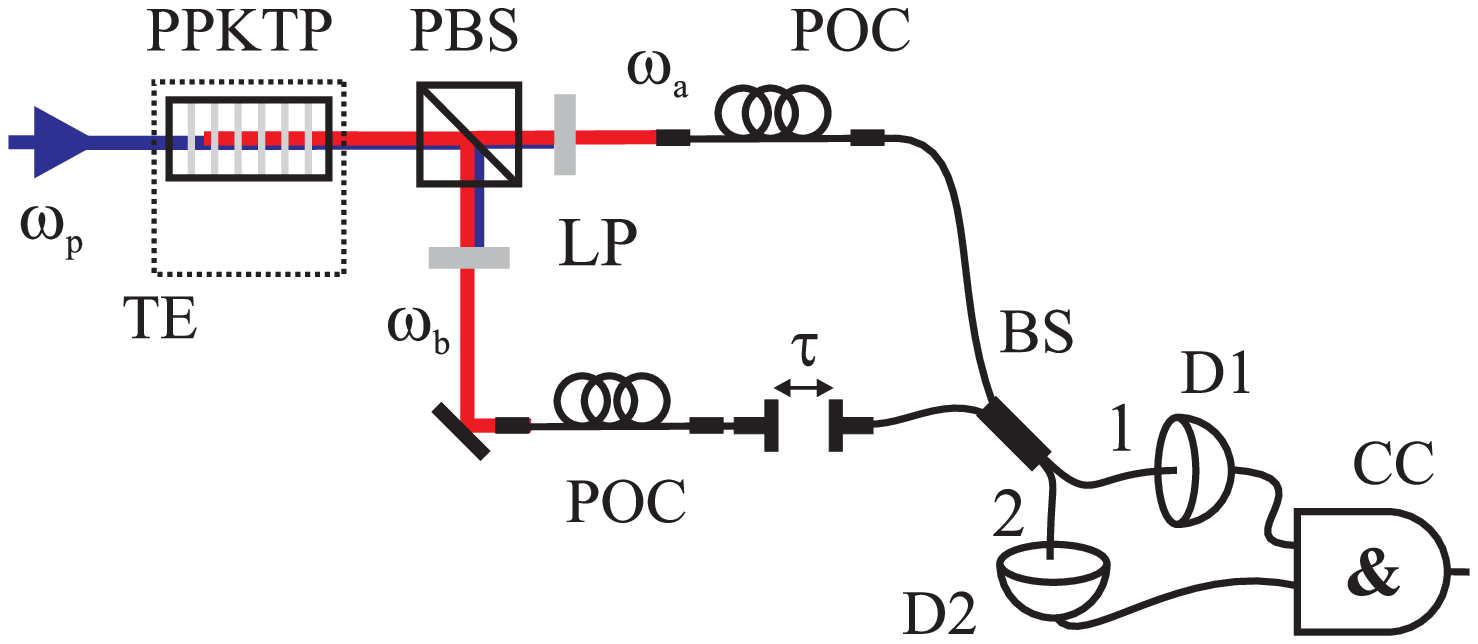}
\caption{Scheme of the experimental setup. Correlated photon pairs were created via parametric downconversion in a nonlinear, type-II quasi-phasematched crystal (PPKTP) which was mounted on a thermoelectric element (TE) and pumped by a $405$~nm cw laser ($\omega_p$). The photons ($\omega_a$, $\omega_b$) were separated on a polarizing beamsplitter (PBS) and coupled into single-mode fibers. The pump light was blocked by long-pass filters. A motorized fiber bridge introduced a variable delay $\tau$ between the photons. Two polarization controllers (POC) guaranteed that the polarization of the photons at the 50/50 fiber beamsplitter (BS) was identical. The events registered by the single-photon detectors (D1, D2) were analyzed by a coincidence counting circuit (CC) within a time window of $4.4$~ns.}
\label{fig:setup}
\end{center}
\end{figure}
Our experimental scheme is shown in figure \ref{fig:setup}. We generated collinear photon pairs with orthogonal polarizations via type-II SPDC in a periodically poled KTiOPO$_4$ (PPKTP) crystal. The temperature $T$ of this crystal was controlled by a thermo-electric element, stabilized to about $0.1\symbol{23}$C. At $T_{deg}=49.2\symbol{23}$C, the photons were emitted at the degenerate wavelength of $810$~nm. As shown in \cite{fedrizzi2007wtf}, the periodic poling allows the photon wavelength to be tuned over a wide range around degeneracy by a change in $T$. The photons of a pair were separated at a polarizing beamsplitter (PBS) and coupled into single-mode fibers. One of the photons passed a fiber air gap which introduced a variable optical delay $\tau$ between the photons before they interfered on a 50/50 fiber beam splitter. Note that in contrast to most previous experiments, we did not use bandpass filters for frequency selection. At the beamsplitter output, the photons were detected by single-photon avalanche photodiodes.

\begin{figure}[!hbtp]
\begin{center}
\includegraphics[width=0.6\textwidth]{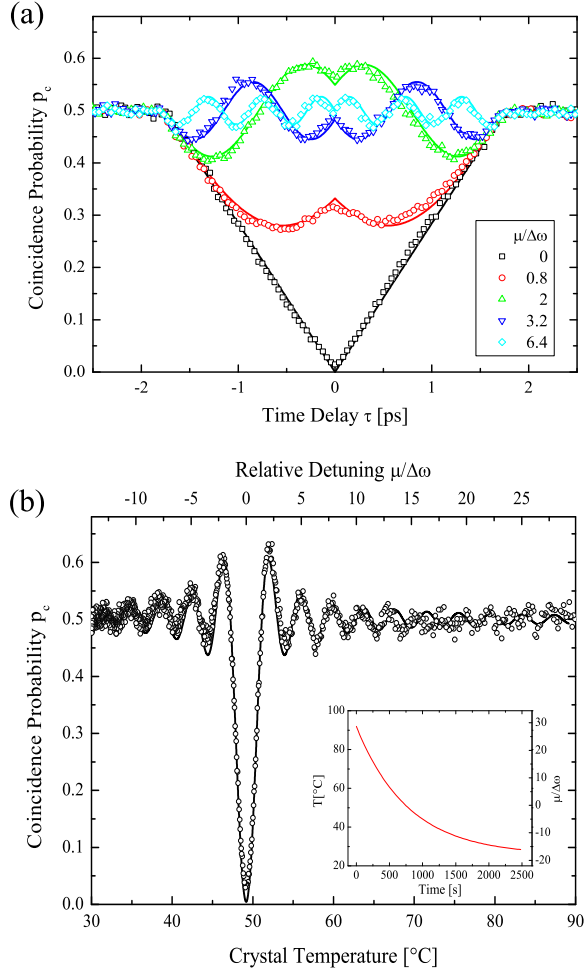}
\caption{(a) Observed coincidence probabilities $p_c$ under variation of delay $\tau$ of the two photons on the BS, at several relative frequency detunings $\mu/\Delta\omega$. (b) Measured $p_c$ vs. crystal temperature $T$ at $\tau=0$. The inset shows the crystal cooling curve and the corresponding frequency detuning $\mu/\Delta\omega$. In both measurements, $p_c$ significantly exceeded the random level of $p_c{=}0.5$, which is a consequence of anti-symmetric frequency entanglement. The solid lines underlying the data in (a) and (b) represent the theoretic predictions from equation \ref{eq:c2general}.} \label{fig:dips}
\end{center}
\end{figure}

This setup allowed us to observe two-photon interference patterns with two continuously tunable parameters; the temporal delay $\tau$ and the photon center-frequency difference $\mu{=}\omega_a^0{-}\omega_b^0$. Figure \ref{fig:dips} shows the measured coincidence probabilities $p_c$ for scans of $\tau$ (a) and $\mu$ (b). The probability $p_c$ is obtained by normalizing the coincidences at the beamsplitter output (no background subtraction) to the rates observed outside the photon coherence lengths.
A scan of the relative temporal delay $\tau$ at zero frequency detuning $\mu{=}0$ yielded a distinct triangular dip, figure~\ref{fig:dips}~(a). The observation of a dip in the coincidence probability was first demonstrated in the famed experiment by Hong, Ou and Mandel (HOM) \cite{hong1987mst} and the triangular shape is characteristic for degenerate photons produced in type-II SPDC \cite{sergienko1995eet,kuzucu2005tpc}. Once we detuned the photon frequencies from degeneracy via discrete changes in the crystal temperature, oscillations emerged in the triangular interference pattern. Eventually, as we increased the detuning $\mu$, the coincidence probability $p_c$ showed harmonic features with a maximum of $p_c^{\mathrm{max}}{=}0.593{\pm}0.002$ which was significantly higher than the random coincidence probability of $p_c{=}0.5$.

The interference pattern as a function of $\mu$ was obtained by fixing the delay at $\tau{=}0$ and heating the PPKTP to $T{=}90\symbol{23}$C, which corresponds to $\mu{=}42.2$~THz. The heating current was then switched off and we recorded coincidences while the crystal cooled down to $28\symbol{23}$C ($\mu{=-}25.4$~THz). The result of this frequency scan is shown in figure \ref{fig:dips}~(b); the inset depicts the crystal cooling curve and the corresponding $\mu$. We clearly observed interference even for detunings far greater than the single photon spectral bandwidth $\Delta\omega{=}1.58$~THz. Again, the coincidence probability periodically exceeded the random level $p_c{=}0.5$. Obviously, in both measurements, frequency-detuned photons showed partial anti-bunching at the BS.

\begin{figure}[htbp]
\begin{center}
\includegraphics[width=0.6\textwidth]{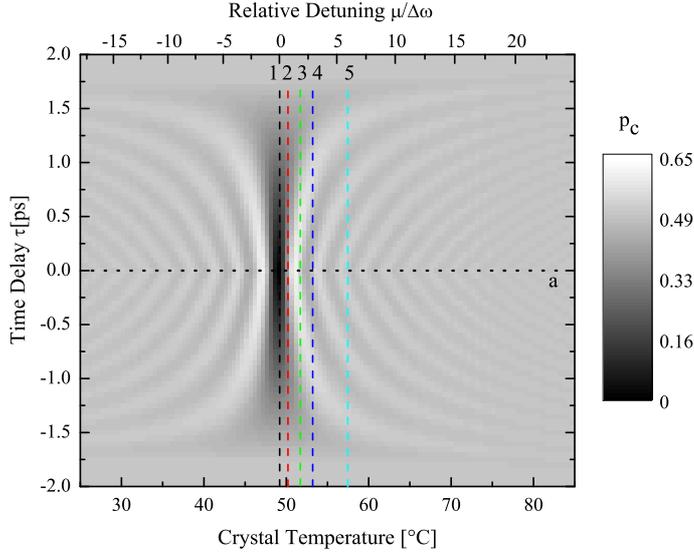}
\caption{Theoretical coincidence probability as a function of optical delay $\tau$ and temperature $T$ (relative frequency detuning $\mu/\Delta\omega$). We experimentally measured the coincidence probability by scanning $\tau$ for different frequency detunings (dashed vertical lines 1-5) and, with variable detuning, for a fixed optical delay $\tau=0$ (dotted horizontal line a).} \label{fig:hommatrix}
\end{center}
\end{figure}

In order to explain our results, we analyze the two-photon state generated by the photon-pair source in figure \ref{fig:setup} theoretically. For collinear type-II phase matching, where the SPDC photons have orthogonal polarization, we obtain the two-photon state \cite{sergienko1995eet}:\begin{equation}\label{eq:state1}
\fl\ket{\psi(\omega_a,\omega_b)}= \int d\omega_ad\omega_b\,
\delta(\omega_p-\omega_a-\omega_b) \mathrm{sinc}\left(\frac{L\Delta
k(T)}{2}\right)a_{a,H}^\dag(\omega_a) a_{b,V}^\dag(\omega_b)\ket{0}.
\end{equation}
Here, $\Delta k(T){=}k_p(\omega_p,T){-}k_a(\omega_a,T){-}k_b(\omega_b,T){-}\frac{2\pi}{\Lambda(T)}$ is the phase mismatch, a function of the optical and thermal properties of the crystal. The energy uncertainty due to the  finite interaction time in the crystal is negligible in regard to the phase matching, and is therefore represented by a $\delta$-function. The spectral amplitude $\mathrm{sinc}(L\Delta k(T)/2)$ emanates from integration of the interacting fields  over the finite crystal length $L$.  We can now expand $\Delta k$ into a Taylor series around ($\omega_a^0$, $\omega_b^0$), where $\omega_a^0$ and $\omega_b^0$ satisfy both energy conservation and phasematching conditions: $\Delta k{=-}(\omega_a{-}\omega_a^0(T))k_a'{-}(\omega_b{-}\omega_b^0(T))k_b'$. Furthermore, we rewrite $\ket{\psi(\omega_a,\omega_b)}$ in terms of frequency differences $\nu{=}\omega_a{-}\omega_b$ and $\mu{=}\omega_a^0(T){-}\omega_b^0(T)$. The spectral amplitude of the two-photon state $\ket{\psi(\mu,\nu)}$ then reads:
\begin{equation}\label{eq:sincnew}
\mathrm{sinc}\left(\frac{L\Delta k(T)}{2}\right)\rightarrow\mathrm{sinc}\left(\frac{\nu-\mu(T)}{\zeta}\right),
\end{equation}
where $\zeta{=}4/(L(k_a'-k_b'))$ is directly connected to the spectral single-photon bandwidth $\Delta\omega$ via $\zeta{=}2\Delta\omega/\pi$.
Next, we introduce the relative optical delay $\tau$ between the two photons of equation \ref{eq:state1} and combine them on a 50/50 beamsplitter. The BS transforms modes $a$ and $b$ into $a_1^\dag(\omega){=}e^{i\omega t_1}/\sqrt{2}(a_b^\dag(\omega){-}i e^{{-}i\omega \tau}a_a^\dag(\omega))$ and $a_2^\dag(\omega){=}e^{i\omega t_2}/\sqrt{2}({-}ie^{i\omega \tau}a_b^\dag(\omega){+}a_a^\dag(\omega))$. At the BS, the photons have identical polarization, so we can neglect the polarization part of the modes. By parametrization of the integral and subsequent calculation (details in Appendix A), we obtain the coincidence detection probability $p_c(\tau,\mu)$ at the two output modes of the BS:
\begin{equation}\label{eq:c2general}
p_c(\tau,\mu)=\cases{
  \frac{1}{2}\left(1-\frac{\sin({\frac{\mu}{\zeta}}(2-\zeta|\tau| ))}{\frac{2
  \mu}{\zeta}}\right) & for $|\tau|< \frac{2}{\zeta}$ \\
  \frac{1}{2} & otherwise.\\}
\end{equation}
We evaluated this expression numerically, using the same set of Sellmeier and thermal expansion equations as in \cite{fedrizzi2007wtf}. The resulting interference pattern as a function of optical delay $\tau$ and frequency detuning $\mu$ is shown in the density plot in figure \ref{fig:hommatrix}. The vertical lines $(1-5)$ and horizontal line $(a)$ mark the parameters along which measurements were taken. The theory curves along these lines are in excellent qualitative agreement to the experimental data in figure \ref{fig:dips} (a) and (b). To reach a good quantitative agreement, we had to scale the theoretic $\mu(T)$ by a factor of $1.25$, which accounts for a discrepancy of the empirical material equations and the actual detuning around degeneracy.

For the degenerate case $\mu{=}0$, (figure~\ref{fig:hommatrix}, line 1) equation~\ref{eq:c2general} correctly reproduces the observed triangle-shaped interference dip, figure~\ref{fig:dips}~(a):
\begin{equation} \label{eq:c2triangle}
p_c(\tau)=\frac{1}{2}\left(1-\wedge\left(\frac{\tau\zeta}{2}\right)\right),
\end{equation}
where $\wedge(x){=}1{-}|x|$ for $|x|{<}1$ and $\wedge(x){=}0$ elsewhere. The base-to-base width of the triangle is $4/\zeta$.

The perfect bunching of the photons at the BS is a consequence of the perfect symmetry of the spectral amplitude of the two-photon state in equation \ref{eq:sincnew} at $\mu{=}0$. For $\mu\neq0$, the spectral amplitude of $\ket{\psi(\mu,\nu)}$ shifts and acquires increasingly anti-symmetric components. Consequently, the perfect interference is diminished and a peak, caused by partial photon anti-bunching, appears at $\tau{=}0$. This peak eventually passes the random coincidence probability of $p_c{=}1/2$.

For the difference frequency dependence at zero delay $\tau{=}0$ (figure \ref{fig:hommatrix}, line $a$), equation~\ref{eq:c2general} reduces to:
\begin{equation} \label{eq:c2sinc}
p_c(\mu)=\frac{1}{2}\left(1-\mathrm{sinc}\left(\frac{2\mu}{\zeta}\right)\right),
\end{equation}
For detuned frequencies, the coincidence probability exhibits damped sinusoidal oscillations which can, in principle, be observed even for arbitrarily large detunings. The highest amount of anti-bunching appears at $p_c(0, \frac{\sim2.25}{\zeta}){=}0.609$, where the amount of anti-symmetry in the spectral amplitude (\ref{eq:sincnew}) reaches a maximum. Because the two-photon state in our experiment is \textit{not} entangled in polarization, this anti-bunching is a direct evidence for entanglement in another degree of freedom---the frequency.

To see that, we first establish the fact that for two interfering, independent photon wave packets, i.e. non-entangled photons with separable spectra, one can never observe a coincidence probability greater than $1/2$ (see Appendix B or \cite{wang2006qtt,eckstein2008bfm}). Observation of $p_c{>}1/2$ is therefore a sufficient, but not necessary, criterion for entanglement: first of all, a 50/50 beamsplitter only acts on the spatial part of a wave function and leaves the spatial symmetry unchanged. In consequence, a spatially anti-symmetric two-particle state has to leave the BS through both output ports, i.e. it anti-bunches, and is therefore certainly entangled. Unfortunately, the spatial part of the wave function of entangled states is not necessarily anti-symmetric but can equally well be symmetric which is the case for the SPDC state in equation \ref{eq:state1} at degeneracy. Here, even though entanglement was present, we observed perfect bunching, which is not a signature of entanglement for bosons. However, as the overall wave function of a bosonic system has to be symmetric, one can always anti-symmetrize the spatial part of a wave function by introducing anti-symmetry in any other degree of freedom. 

Using this method, we were able to reveal the underlying frequency entanglement of the SPDC photons by appropriate frequency detuning. As the photons in our experimental configuration were confined to transversal single modes and the polarization modes were not affected by a change in frequency, the introduced spectral anti-symmetry in the term (\ref{eq:sincnew}) therefore had to cause spatial anti-symmetry, which was detected at the BS.

Note that, instead of changing the photon center frequencies, it is also possible to indirectly influence the spectral properties of the two-photon state, e.g. by introducing dispersive elements into the photon paths \cite{okamoto2006ttp,brida2006dsb}. Also, one can tune entirely different degrees of freedom to achieve spatial anti-symmetry, e.g. the transversal part of the wave-function in multi-mode HOM interferometry \cite{nogueira2001eos,walborn2003mhm}. Two well known examples for states where the entangled degree of freedom is very easily accessible are path-entangled \cite{michler1996ibs} and polarization-entangled Bell states \cite{mattle1996,kim2005qid}. Here, one can change the spatial symmetry of the states with a simple (birefringent) phase shifter.

We further remark that a similar phenomenon should arise for the case of two fermions incident on a beamsplitter, as for example in electron interferometry \cite{neder2007ibt,oliver1999hba}. The total state of two fermions must be anti-symmetric and two independent and indistinguishable fermions in two spatial modes would therefore anti-bunch \cite{lim2005ghm}. In this case, \emph{bunching} would be a clear signature of entanglement.

In conclusion, we presented the first interference-filter free, tunable, single-mode spatial quantum beating experiment. This is a prerequisite for a number of interesting applications in quantum information, e.g. for the preparation of discrete, tunable color entanglement \cite{ramelow2009dtc}. We observed frequency entanglement of photons generated in SPDC without actually performing frequency correlation measurements, which would have been hard to implement. We achieved this by anti-symmetrization of the initially symmetric spatial part of the wave function and observation of two photon interference on a simple beamsplitter to reveal the spatial anti-symmetry of the states which is a sufficient criterion for entanglement. We propose to use this method for the detection of hidden entanglement in general fermionic and bosonic quantum systems.

\section*{Acknowledgments}
Thanks are due to J. Kofler, T. Paterek , T. C. Ralph, A. G. White and M. \.Zukowski for valuable discussions. This work was supported by the Austrian Science Fund (FWF), by the Austrian Research Promotion Agency (FFG) and the EC funded project QAP. M. B. acknowledges support by the Australian Research Council and the IARPA-funded U.S. Army Research Office Contract No. W911NF-05-0397. T. J. acknowledges support by the Australian Research Council (Linkage International Program).

\appendix
\setcounter{section}{1}
\section*{Appendix A} We start with the SPDC state in
equation \ref{eq:state1} that is transformed at the BS. Parametrization of the integral and projection onto output modes $1, 2$ yields the probability amplitude $A(t;\tau)$ for a coincidence in $1$ and $2$:
\begin{equation}
A(t;\tau)=\frac{1}{2}\int d\nu
\,\mathrm{sinc}\left(\frac{\nu-\mu}{\zeta}\right)\left(e^{-i\nu
t}-e^{i\nu (t+\tau)}\right),
\end{equation}
where $t=(t_1-t_2)/2$. This integral can be solved via a Fourier transform, and gives:
\begin{equation}
A(t;\tau)=\frac{1}{2}\left( e^{i\mu t}\Pi\left(\frac{
t\,\zeta}{2}\right)-e^{-i\mu(t+\tau)}\Pi\left(\frac{(
t+\tau)\,\zeta}{2}\right) \right),
\end{equation}
where $\Pi(x)$ is the rectangular function $\Pi(x)=1$ if
$|x|\leq1/2$, and $\Pi(x)=0$ elsewhere. Thus, the coincidence detection probability in the output modes is:
\begin{eqnarray}
\fl p_c(\tau)=A_0\int dt\,|A(t;\tau)|^2=\frac{A_0}{2}\biggl(\int dt\,
\Pi\left(\frac{t \,\zeta}{2}\right)-\nonumber\\-Re \int dt\,
e^{i\mu(2t+\tau)}\Pi\left(\frac{t\,\zeta}{2}\right)\Pi\left(\frac{(
t+\tau)\,\zeta}{2}\right)\biggr).\label{eq:integral3}
\end{eqnarray}
By evaluating the first integral of this expression, we find the normalization constant $A_0=\frac{\zeta}{2}$. Evaluation of equation \ref{eq:integral3} then leads to $p_c(\tau)$ in equation \ref{eq:c2general}.

\section*{Appendix B}
We now consider a separable state of two photon wave packets with identical polarization, produced by independent sources. A generalized, separable state is:
\begin{equation}
\ket{\psi[f,g]}=\int d\omega_1d\omega_2\, f(\omega_1)g(\omega_2)\,
a^\dag(\omega_1)b^\dag(\omega_2)\ket{0},
\end{equation}
where $f(\omega_1)$ and $g(\omega_2)$ are properly normalized spectral amplitudes. Some calculation leads to:
\begin{equation}
\label{separable}
p_c=\frac{1}{2}-\frac{1}{2}|\tilde g*\tilde f(\tau)|^2.
\end{equation}
It is obvious that $p_c$ for separable states is always less than $1/2$. For mixed states, one observes incoherent contributions in the form (\ref{separable}), from which no interference effects arise. Thus, also in the case of a mixed state, we expect $p_c<1/2$.

\section*{References}
\bibliographystyle{unsrt}

\end{document}